\documentclass[graybox]{svmult}

\usepackage{mathptmx} % selects Times Roman as basic font
\usepackage{helvet} % selects Helvetica as sans-serif font
\usepackage{courier} % selects Courier as typewriter font
\usepackage{type1cm} % activate if the above 3 fonts are
\usepackage{mathtools} 		% spezielle Symbole
\usepackage{amssymb,amsmath,bbold}
\usepackage{makeidx} % allows index generation
\usepackage{graphicx} % standard LaTeX graphics tool
\usepackage{multicol} % used for the two-column index
\usepackage[bottom]{footmisc}% places footnotes at page bottom

\newcommand{\req}[1]{Eq.\,(\ref{#1})}

\newcommand{\rf}[1]{Fig.~\ref{#1}}

\makeindex % used for the subject index
\begin{document}

\title*{Die Erste Stunde}
\author{Johann Rafelski}
\institute{Johann Rafelski \at Department of Physics, The University of Arizona, Tucson, AZ 85721\\ \email{Rafelski@Physics.Arizona.EDU}\\{\it Presented at:}
 FIAS International Symposium on {\it Discoveries at the Frontiers of (Walter Greiner) Science}, Frankfurt, June 26 - 30, 2017. to appear in proceedings
%\and Name of Second Author \at Name, Address of Institute \email{name@email.address}
}
\maketitle

\abstract{I recall my \lq\lq first hour\rq\rq\ events following on my meeting in Fall 1968 in the classroom with my academic teacher and thesis mentor Prof. Dr. Dr. h.c. multiple Walter Greiner. My comments focus on the creation of the new \lq\lq strong fields\rq\rq\ domain of physics in Frankfurt. I argue that this was the research field closest to Walter\rq s heart during his lustrous academic career. I will describe the events that lead on to Greiner\rq s course books, Walters actions leading to the rise of Frankfurt School of Theoretical Physics, and show how a stability principle defined his science.}

%%%%%%%%%%%%%%%%%%%%%%%%%%%%%%%%%%%%%%%%%%%%%%%%
\section{Introduction}
\label{sec:intro}

Walter Greiner arrived in Frankfurt in the mid 1960s. He came in as a reformer, pushing through many changes at the Physics Faculty (Fachbereich Physik) of the Johann Wolfgang von Goethe Universit\"at in Frankfurt. In Fall 1968, Walter\rq s newly approved Theoretical Physics course was offered to physics students in their first semester and attracted many students, including the author. For me and many others in this class the meeting with Walter was a random chance. However, we stayed on because of Walter. 

Teaching freshman Theoretical Physics Course was an educational revolution. It was accompanied by another revolution; West Germany was in the midst of a large scale student revolt. I recall that the J.W. Goethe University could not set exams; some courses were even canceled as the zealots focused on particular \lq reactionary\rq\ lecturers. Other courses were disrupted temporarily by sit-ins organized by idealistic students responding to the crooked but active Soviet propaganda machine operating from East Germany.

A reader interested in a more general characterization of life and work of Walter Greiner should consult another recent commentary~\cite{Biro:2018wyo}. This article is very different as it describes in form of personal reminiscences an important series of events that occurred mainly between 1968-1983 during which 15 year period I interacted strongly with Walter. 

%%%%%%%%%%%%%%%%%%%%%%%%%%%%%%%%%%%%%%%%%%%%%%%%
\subsection{Frankfurt 1968-1971}
In Greiner\rq s classroom nobody from within disrupted the lectures. Moreover, when external non-physics revolutionaries tried to stop Greiner\rq s Theoretical Physics freshman class, his students defended the classroom, expelling the non-course students, throwing back the stink bombs and barricading the entrance doors from within with chairs and desks. Despite numerous distractions, and the absence of formal examination (for fear of external disruption), there was lots of learning going on. We had classes, regular tutorial study groups, and the teaching program proceeded well. Among the 70 or more freshman students, many made great scientific careers. 

I think that this shows that teaching in the challenging way Walter pioneered in Frankfurt leads to success irrespective of situation. And for those going on to academic careers the near complete lack of formal examination may have been an asset: learning was not distracted by examination. Of the 1968 crop of students, quite a few joined later Greiner research group, and several became tenured theoretical physics full professors.

This concentration of talent around Walter was due to his proactive approach: Walter cared for and developed young men and women working with him. Walter was bent on keeping his talented students. When someone made a move that displeased Walter\rq s vision, he would straighten out the situation, typically offering his assessment on who was who in theoretical physics. His decisive and convincing arguments were without doubt an important reason for the successful birth of the Frankfurt School of Theoretical Physics in the eventful months of 1968-70.

I still remember how Walter reacted when I told him that I won a very coveted and competitive Studienstiftung fellowship to Oxford. Walter was upfront and direct: \lq\lq ... only a!\@\#s are working there.\rq\rq\ Also, within days of this conversation, shortly after I completed my Diploma, and despite being only 21 years old, I was appointed his Scientific Assistant, see \rf{JRWiss}.

%
%%%%%%%%%%%%%%%%%%%%%%%%%%%%%%%%%%%%%%%%%%%%%%%%
\begin{figure}[t]
\sidecaption%[t]
%align sidecaption with the top of the figure optional argument [t]
\centerline{
\includegraphics[width=0.9\columnwidth]{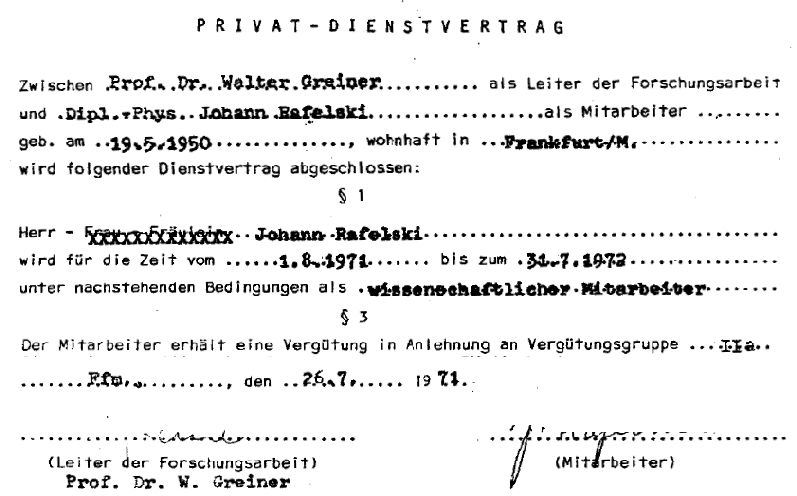}}
%\includegraphics[scale=.65]{./WG70.jpg}
%\picplace{5cm}{2cm} % Give the correct figure height and width in cm
%
\caption{Walter Greiner appoints the author as his \lq\lq Scientific Assistant\rq\rq\ in July 1971 {\it Source: Johann Rafelski archives}}
\label{JRWiss}
\end{figure}
%%%%%%%%%%%%%%%%%%%%%%%%%%%%%%%%%%%%%%%%%%%%%%%%%%%%%%%

A few \lq older\r\ (as compared to me) Assistants supported Walter\rq s teaching and research efforts. Burkhard Fricke, and Ulrich Mosel come to my mind, Burkhard was leading my tutorial group. Ulrich was the primary pillar of classical Nuclear Science and did both his Diploma and PhD with Walter. However, I was Walter\rq s first hire from among the people he taught from the first semester on, beginning in Fall 1968. I in turn introduced Walter to others who became important in the strong field physics formative years, including Berndt M\"uller, Gerhard Soff, and a year later Joachim Reinhardt, the future soul of Walter\rq s rapidly expanding strong field research group.

%%%%%%%%%%%%%%%%%%%%%%%%%%%%%%%%%
\subsection{Theoretical physics course books}
It is of interest to many to understand how and why Walter Greiner\rq s red-book series, \lq\lq Theoretische Physik,\rq\rq\ was created. Before Greiner, nobody dared to teach a Theoretical Physics course in the 1st semester; thus no appropriate textbooks were available for freshman student use in the study of theoretical physics. In order to teach material which previously would appear typically only two years later in the curriculum, Walter clearly recognized the need to simplify, to explain by example, and to offer full solutions of exercises.

To create such a new series for students, Walter realized that much of it had to be co-written by students. His \lq\lq Assistant,\rq\rq\ Burkhard Fricke who set the tutorial exercises (\"Ubungen) for each week, would also collect from student volunteers their class protocols of class lectures, exercises, and solved problems. These were reviewed and edited by Burkhard. Walter would make another set of edits before these notes were typed, figures drafted and all put together into a \lq\lq Script\rq\rq\ by Walter\rq s secretarial staff, supervised by the \lq\lq Assistant\rq\rq. 

The firs class course \lq\lq Scripts\rq\rq\ of Classical Mechanics I+II and Electrodynamics were prepared by Burkhard Fricke, as can be read in \rf{WGEMbook}. Scripts were printed just like a PhD thesis by an in-house printing press maintained by the Theoretical Physics Institute. These scripts were available to anyone interested and were popular among students selling at 2\,DM a copy, the cost of a lunch in a student Mensa restaurant. 

Walter would use the first edition of his script, repeating three years later the same series of classes. Doing this he caught inevitable errors, and added and expanded the material. A second edition of the script was then created following the path of the first. In the case of {\it Electrodynamik} shown in \rf{WGEMbook}, the first script edition was in late 1970; the second in late 1973 and the book version was ready to go to press a year later. 
 
%%%%%%%%%%%%%%%%%%%%%%%%%%%%%%%%%%%%%%%%%%%%%%%%
\begin{figure}[t] 
\sidecaption%[t]
%align sidecaption with the top of the figure optional argument [t]
\centerline{%
\includegraphics[width=0.9\columnwidth]{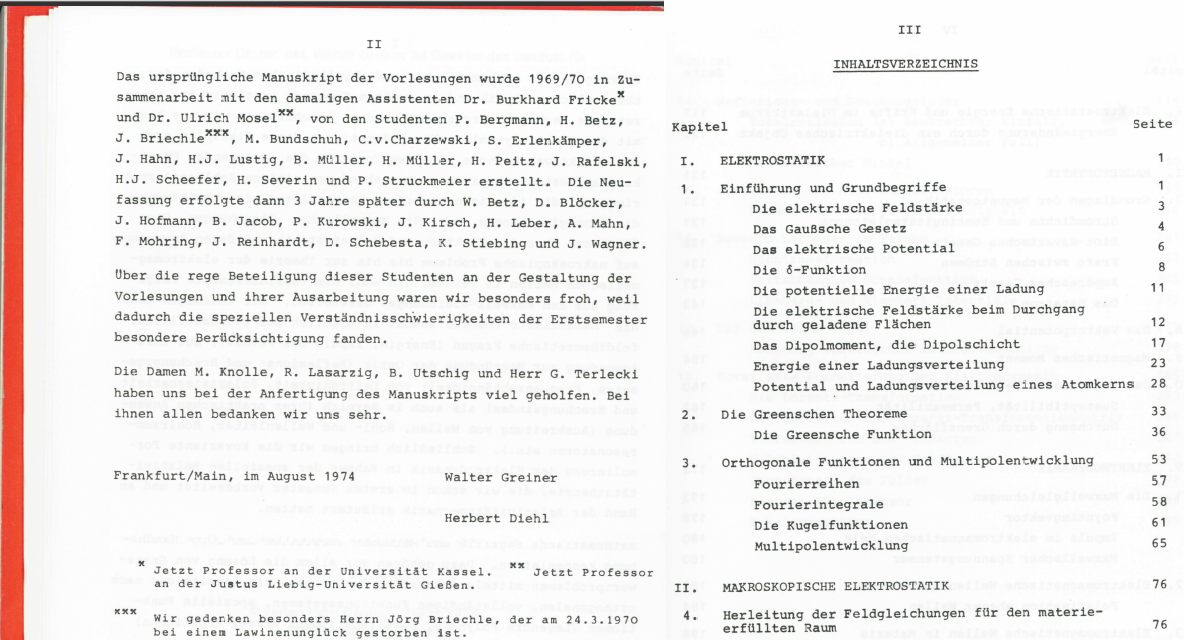}}
%\includegraphics[scale=.65]{./WG70.jpg}
%\picplace{5cm}{2cm} % Give the correct figure height and width in cm
%
\caption{View on introductory remarks in the W. Greiner and H. Diehl {\it Electrodynamik} (Verlag Harri Deutsch, Frankfurt 1975, ISBN 3871441856). See text for further comments {\it Source: Johann Rafelski archives}}
\label{WGEMbook}
\end{figure}
%%%%%%%%%%%%%%%%%%%%%%%%%%%%%%%%%%%%%%%%%%%%%%%%%%%%%%%

As the fame spread, a publisher, Verlag Harri Deutsch, located just across the street from the institute became interested. So the 2nd edition of the lecture script was published reprinted directly from the script document. In this way Greiner\rq s script became available beyond Greiner\rq s institute. This in turn opened the path for other German Universities to embark on Greiner\rq s teaching method. The newly developed Theoretical Physics teaching model began to influence all German-speaking Universities.

I selected the pages of the {\it Electrodynamik} 1st book edition in \rf{WGEMbook} to show that Walter credited everyone deserving for the help in creating the main course books: we can read the names of all involved students and Assistants, and Herbert Diehl of 2nd Script edition is coauthor. After 40 years one would think that most if not all students listed in this page are long forgotten. This is not the case: many embarked on great careers, some returned (Johannes Kirsch for example) after their retirement contributing to the academic life in the newly formed Institute for Advanced Studies.

Aside of the core stream of books Walter saw the need and also an opportunity for additional texts offering specialized topics. While I was teaching Special Relativity in 1981, Walter asked me if I wanted to contribute as an author to the new topical series he would publish as \lq\lq Herausgeber\rq\rq. I agreed and Walter was keen to see my text as soon as possible; it was to be the first in the topical series. Being a young professor I did not at first have an \lq\lq Assistant\rq\rq\ to help me and I was teaching without sources as secondary literature was inaccurate and misleading. Moreover, I had research priorities. As a result three years passed before the {\it Spezielle Relativit\"atstheorie} book was ready~\cite{SR1}. 

By that time the original publication model for the special topics series was colored by other projects and the question of economic success: Walter Greiner was a successful author, Johann Rafelski (and the authors of other topical projects underway) were not. Verlag Harri Deutsch wanted to see the special series books as Greiner\rq s books. To account for the fact that I did all of the author\rq s work, Walter proposed to give me 3/4 of author compensation, considering that editors of series in general get some fraction of the revenue. Indeed some 15 years later when publishing a book on Relativistic Heavy Ion physics, I was offered the same contract with Cambridge University Press; the difference was that when the CUP book appeared, the authors, and series editors, were clearly set apart. This was not the case with Verlag Harri Deutsch.

My German relativity text was very successful: in 10 years and three editions it sold 6,300 copies. The 2nd and 3rd editions where hard cover bound. However, the used physics book market trades today soft cover reprints of the 3rd edition. Walter used to say that one way to test popularity is to discover that one\rq s books are stolen in libraries. Together we made it to a higher dimension of popularity: the German Special Relativity book is so popular that someone prints softcover wild copies.

Several of Walter\rq s books have been published in English by Springer Verlag (beginning in 1989). Seeing the success of \lq\lq Relativity\rq\rq\ and the fact that after 1983 I was teaching in English we discussed a possible English edition. However, Walter, I, Verlag Harri Deutsch, not to mention Springer Verlag, never found a contractual solution. Our book went out of press by 1995/6. Finally, after 20 more years, in 2017, I published a new English book on special relativity~\cite{Rafelski:2017hyt}, dedicating this work to Walter who had passed away a few months earlier.

%%%%%%%%%%%%%%%%%%%%%%%%%%%%%%%%%%
\subsection{The beginning of strong field physics in Frankfurt}
Following on my \lq\lq Vordiplom\rq\rq\ in early 1970 I was drawn into a new topic of \lq\lq strong fields\rq\rq. Walter correctly saw that I would best fit this intellectual adventure that even today seeks an intellectual home, falling often through the cracks that open up between particle, nuclear, atomic foundational physics domains.
 
Walter needed someone ready to jump in since by mid-1969 he had convinced himself that we had not understood the physics of atomic electrons bound to superheavy large atomic number $Z$ nuclei. It is best here to let Walter speak, citing from the conference panel discussion printed in the proceedings of the International Conference on Properties of Nuclear States held in Montreal, Canada August 25-30, 1969. This was the premier meeting event where who-was-who in the international nuclear science appeared. 

We read (page 611 in \cite{WG1969}): \lq\lq {\it Greiner:} The important thing is that for $Z=80$ you have $Z\alpha$ ($\alpha\simeq 1/137$ (JR)) less than unity, but for super-heavy nuclei around $Z=164$ it is suddenly larger than unity and you do not know whether the expansion in $Z\alpha$ converges anymore. You really have to start from a completely different point of view and develop new methods.\rq\rq\ Responding and addressing challenges by G.E. Brown and D.H. Wilkinson, Walter continues (page 612): \lq\lq \ldots the $1s$-electron levels are very quickly very strongly bound and dive into the lower continuum. Their binding energy very quickly increases up to 1\,MeV. This is the point where the difficulties arise. \ldots I would like to stress that this quantum electrodynamics problem is very interesting from a purely theoretical point of view.\rq\rq\ 

The last remark shows what attracted Walter to the topic, while the first part merely explains how he came to consider the research program in the first place. Walter knew that the island of nuclear stability at $Z=164$ would not lead to what we call today supercritical fields, but of course this did not matter to him and as the following developments showed, we had an alternate path, the quasi molecules. More on this later; see Sect.\,\ref{QMol}. Continuing the timeline: already in August 1968 Walter and W. Pieper submitted to Z. Physik results showing the need for $Z>172$ \cite{69Pieper}. This publication was delayed while a partial redo of the computation was carried out~\cite{69Rein}, in which an editor of the journal was thanked for suggesting the research topic. 

The results by Pieper and Greiner confirmed and quantified using realistic nuclear charge distributions the earlier results obtained by Werner and Wheeler~\cite{58WW}, who in their 1958 publication abstract say: \lq\lq Despite $Z$ values substantially higher than 137, the $K$ electrons behave perfectly normally because of the finite extension of the nucleus. Vacuum polarization and vacuum fluctuations are roughly estimated to make relatively minor alterations in the $K$ electron binding -- which exceeds $mc^2$.\rq\rq 

The above describes succinctly the state of knowledge before Pieper and Greiner: what Werner and Wheeler {\it overlooked} is that at sufficiently large $Z>172$ the problem that earlier was seen for $Z>137$ reappears and does so in a way that is even less comprehensible. While for the point source of the electromagnetic field the Dirac Hamiltonian becomes non-selfadjoint, meaning that the spectrum of bound states is not complete, for a finite nuclear size when the electron biding exceeds $E>2mc^2$ a bound state \lq\lq dives\rq\rq\ into the antiparticle solutions of the Dirac equation. While some work (incorrectly) claimed that the old self-adjointness problem related to the $1/r$ singular potential returns, Walter never made this mistake.

I have little doubt that in order to make progress someone as unencumbered by prior thinking as Walter had to become interested in strong binding, who also had to be a person with a wide knowledge of diverse theoretical tools and methods. In particular Walter was aware of the work by U. Fano on embedding of bound states in a continuum~\cite{Fano1961}. This created the basis for the understanding of the positron autoionization phenomenon, see Sect.\,\ref{FanoD}.

As we will describe in Sect.\,\ref{BIModel}, the study of nonlinear limiting field electromagnetism paved the way to the recognition that even in a limiting force theory, there is no way to avoid the phenomenon of electron binding in excess of $2mc^2=1.022$\,MeV. In colloquial language we call this level crossing into negative energy continuum \lq\lq diving\rq\rq. This behavior had been described and left unresolved in earlier studies.

Looking back I remember how in the Winter 1971/72, the strong field group met regularly Saturday mornings in Walter\rq s office suite located in the SW corner of the 5th floor of the Physics building at Robert-Mayer Strasse 10. One Saturday morning, in an spontaneous burst of creativity, Walter adapted Fano\rq s renown work to the case of \lq\lq diving\rq\rq, calling this process autoionization of positrons, implying that one should interpret \lq\lq diving\rq\rq\ of particle states into antiparticle continuum as a phenomenon where a hole, a vacancy in the bound state turned in the \lq\lq diving\rq\rq\ process into a spontaneously emitted positron; we return to this topic in Sect.\,\ref{FanoD}.

I believe that to the end of his life Walter considered this insight his greatest. He never hesitated to tell about this process. For example, at the \lq\lq International Conference on Strangeness in Quark Matter\rq\rq\ (SQM2006), held at the University of California at Los Angeles (UCLA), Walter Greiner gave the UCLA Departmental colloquium, \lq\lq On the Extension of the Periodic System into the Sector of Strange- and Antimatter.\rq\rq\ 

In this UCLA lecture Walter described to a large and multidisciplinary audience the main domains of research he developed in Frankfurt, beginning in the late 1960s. As the lecture unfolded we sensed Walter\rq s heart beating loudest when he recounted the strong field QED. Water described in detail the physics of \lq\lq diving\rq\rq\ seen in \rf{fig:SQMDiving}, as if this was still the Spring 1972 when we published this result together~\cite{Muller:1972zza}. 
 
%%%%%%%%%%%%%%%%%%%%%%%%%%%%%%%%%%%%%%%%%%%%%%%%
\begin{figure}[t] 
%\sidecaption%[t]
\centerline{%
\includegraphics[width=0.9\columnwidth]{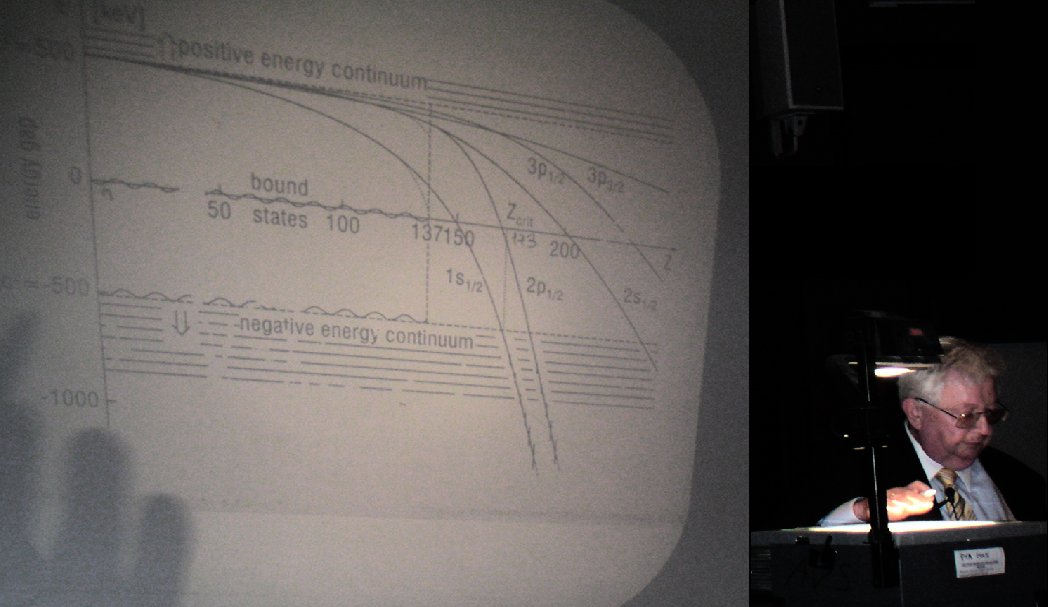}}
\caption{Walter Greiner showing in March 2006 a slide created with contents created for publication in Spring 1972~\cite{Muller:1972zza}. {\it Photo: Johann Rafelski}}
\label{fig:SQMDiving} 
\end{figure}
%%%%%%%%%%%%%%%%%%%%%%%%%%%%%%%%%%%%%%%%%%%%%%%%%%%%%%%

%%%%%%%%%%%%%%%%%
\subsection{The stability principle}

Even though in science dogmatic principles are rarely of value, as the context and ultimate fate of the false Aristotelian Physics proves by example, Walter's adherence to a \lq\rq stability principle \rq\rq\ guided, and may have misdirected some of his effort. Let me show in a few examples how this worked.

\paragraph*{Strong fields}
Walter was not willing to embrace his own Pieper-Greiner result unconditionally. In 1970, when he offered me a Diploma Thesis topic, he was searching for a new mechanism to make the critical binding go away. He was searching for improvements such that in the presence of ever larger externally applied fields, the binding energy of an electron should never reach the limit $E\to 2mc^2\simeq 1$\,MeV. He followed several paths of which one was the modification of the electromagnetism, which I will describe in technical detail in Sect.~\ref{BIModel}.

We found that other experimental results limited the opportunity to modify electromagnetism and thus we concluded that there was no chance that this approach could be of relevance for stabilizing the Dirac equation solutions and resolving the electron \lq\lq diving\rq\rq\ problem, see Sect.~\ref{FanoD}. Thus in the pivotal 1972 publication where we quantitatively demonstrate the positron instability, Walter refers to this situation as follows~\cite{Muller:1972zza}: \lq\lq At this point the single-particle theory seems to break down. Pieper and Greiner and later Popov have interpreted this to mean that electron-positron pairs are created spontaneously.\rq\rq\ The word \lq\lq seems\rq\rq\ reflects on his continued hope that that non-perturbative QED many body theory effect could be significant.

I think Walter's change of heart in regard to \lq\lq diving\rq\rq\ behavior occurred with our 1973 work on \lq\lq Charged Vacuum\rq\rq~\cite{Rafelski:1974rh}. The reinterpretation of the phenomena in terms of vacuum structure and the insight that the structured quantum vacuum acquires localized charge density overwhelmed Walter's adherence to stability. 

Into this new context arrived Miklos Gyulassy, freshly minted at Berkeley. Miklos told me, and I agree, that his Frankfurt job and excellent relationship with Walter was a direct outcome of him showing independently of our effort (in a very elaborate numerical work~\cite{Gyulassy:1974ba}) that the Frankfurt Charged Vacuum theory was right.

\paragraph*{Event horizon}
Walter posited that gravity should not prevent light from traveling; in other words, an event horizon associated with a black hole solution should not exist. As strong field physics advanced, Walter saw the connection to gravity, and he hoped that there would be a way, using the ideas we developed in relativistic quantum theory of strong fields, to modify Einstein\rq s gravity. He gained this insight in his very first General Relativity (GR) class, which I recall he held in Winter 1971/2. Significant effort went into the understanding of Dirac equation solutions in a strong gravity field~\cite{Soffel:1980kx}. Following in the footsteps of related work on nonlinear EM theory, see Sect.~\ref{BIModel}, students in Frankfurt worked on what we call today $f(R)$ gravity~\cite{Muller:1978zz}, where action is a nonlinear function of the Ricci tensor $R$. 

When I visited the GSI laboratory in Summer 1977 for three months on the way from US to CERN, Walter wanted me to create a no-event-horizon gravity, asking for my full commitment. In these three months I learned more about gravity than I did in the rest of my life. Aside of me, Berndt M\"uller became also part of effort. We did not solve the problem, and I must add, Berndt and I challenged Walter, if the existence of an event horizon was really a problem? I believe it was the first time that a clear divisive line opened between Walter and his first hour students. It is worth noting that Walter never relented about black holes, and after some 30 more years he published a no-event-horizon gravity theory~\cite{Hess:2008wd,HessWG2016}.

Even if our preoccupation with event horizon did not lead to a good outcome, our effort paid off in a different way. Berndt M\"uller, Walter and I published on the interpretation of strong external EM field -- thus acceleration -- in terms of effective temperature~\cite{Muller:1977mm}. This work implies that some deep connection exists between EM and GR and has influenced my work from the past decade~\cite{Labun:2012jf}.

%\subsubsection*
\paragraph*{Quark-gluon plasma}
The development of the GSI laboratory near Frankfurt in Wixhausen, now part of the City of Darmstadt, was driven by the hard work and political skill Walter so often displayed. This laboratory today is a renowned center of relativistic heavy ion research. Among the most important physics developments that occurred in late 1970s and early 1980s was the exploration of nuclear matter using beam of relativistic heavy ions. Walter was the pioneer in this field, working with Horst St\"ocker, another lustrous student and recent GSI director, on shock waves that nuclear matter should support, see for example~\cite{Stocker:1981zz}. 

This work began in close cooperation with Erwin Schopper, the founder of (experimental) \lq\lq Institut f\"ur Kernphysik\rq\rq\ (nuclear physics). For this work The European Physical Society in 2008 awarded to Walter Greiner and to Schopper\rq s successor, Reinhard Stock, the {\bf Lise Meitner Prize} for nuclear science. Walter\rq s citation reads: \lq\lq \ldots for his outstanding contributions to the development of the field of relativistic nucleus-nucleus collisions by pioneering the ideas of shock waves and collective flow in nuclear matter, thus inspiring experimental studies of nuclear matter at extreme conditions of density and temperature.\rq\rq 

There is no mention in this citation of the new state of matter, the deconfined quark-gluon plasma (QGP) phase of matter which had perhaps even more consequential impact on nuclear physics. By the late 1970s the recognition grew that this deconfined form of matter could be created in ultra relativistic heavy ion collisions. I reported on the early work on deconfinement, and QGP formation, in the \lq\lq Hagedorn\rq\rq\ volume~\cite{Rafelski:2016hnq}.

The QGP research direction fit both GSI and Greiner\rq s traditional nuclear research program perfectly. He and his group should have been among the pioneers in this new research field. However, in Frankfurt the stability concerns affected the early development of the QGP physics: the idea that nucleons at high temperature could melt and dissolve into the more fundamental constituents of matter, quarks and gluons, did not sit easily in Walter\rq s mind.

\begin{figure}[t] 
%\sidecaption%[t]
\centering
%align sidecaption with the top of the figure optional argument [t]
\includegraphics[width=0.9\columnwidth]{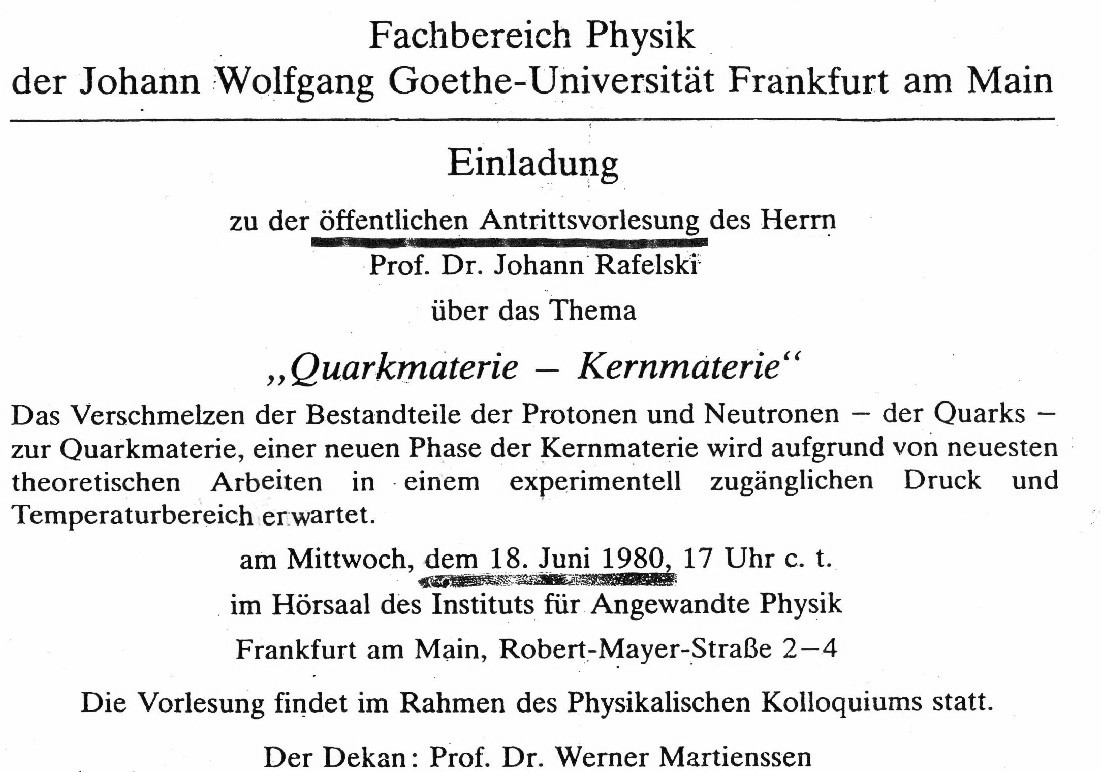}
%\includegraphics[scale=.65]{./WG70.jpg}
%\picplace{5cm}{2cm} % Give the correct figure height and width in cm
%
\caption{The translation of the abstract of author\rq s public Inaugural lecture held in Frankfurt on June 18, 1980 reads: {\it The fusion of constituents of protons and neutrons -- quarks -- into quark matter, a new phase of nuclear matter has been studied in recent theoretical work. It is expected to occur in an experimentally accessible domain of temperature and pressure.\rq\rq\ Source: Johann Rafelski archives}}
\label{JRInaugural}
\end{figure}

Walter attended my inaugural lecture event on June 18, 1980, see Fig.\,\ref{JRInaugural}, where a contemporary view of these QGP developments was offered based on work I had done at CERN in the prior 2.5 years. I hoped and expected my friend and teacher to sit in the front row nodding approvingly. Instead, he was in the very back of the filled room, not listening as I observed with some trepidation. 

In following years Walter continued in clear and outspoken opposition; several PhD students in his group were working to prove that QGP could not be observable. I departed Frankfurt in 1982, heading back to CERN and later on to Cape Town. Despite these setbacks, in the ensuing years the QGP effort in Frankfurt grew stronger around some of my students who persevered -- Peter Koch deserves to be mentioned and praised for this effort; see for example Ref.\,~\cite{Greiner:1987tg}.

%
%%%%%%%%%%%%%%%%%%%%%%%%%%%%%%%%%%%%%%%%%%%%%%%%
%%%%%%%%%%%%%%%%%%%%%%%%%%%%%%%%%%%%%%%%%%%%%%%%%%%%%%%

%%%%%%%%%%%%%%%%%%%%%%%%%%%%%%%%%%%%%%%%%%%%%%%%%%%%%%%%%
\section{Born-Infeld Nonlinear Electromagnetism}\label{BIModel}
Following Max Born's passing in January 1970, his work on nonlinear Born-Infeld (BI) electromagnetism was widely discussed. Walter Greiner, given his adherence to stability, was fascinated by the BI effort to stabilize electromagnetism. In the BI-EM theory the self-energy of a charged point particle was finite and the acceleration just like particle speed had a maximum value. 

Walter believed that if the electric field of an atomic nucleus had a limit, there could be a major change in the solutions of the Dirac equation and the critical binding behavior might disappear. This was the project he signed me on for my Diploma, and arranged for assistance from Lewis P. Fulcher. Lewis had just arrived in Frankfurt as a postdoctoral fellow, having graduated with Judah M. Eisenberg from the University of Virginia.

To understand the working of the BI-EM, we need to truly distinguish between EM field and the displacement fields. We note the inhomogeneous Maxwell\rq s equations
\begin{equation}\label{LF2}
\displaystyle
\vec \nabla\cdot \vec D=e\rho\;, \quad 
\vec \nabla\times \vec H= \displaystyle\frac{\partial D}{\partial t}+\vec ej\;,
\end{equation} 
connecting with $\vec E$ and $\vec B$ by means of the first derivative of the Lagrangian density 
\begin{equation}\label{BI2bis}
\vec D=\displaystyle \frac{\partial \cal{L}}{\partial \vec E}\;,\qquad
\vec H=\displaystyle \frac{\partial \cal{L}}{\partial \vec B} \;.
\end{equation} 
This is well known to those who practice electromagnetism in media. Max Born worked with a covariant medium by choosing a nonlinear covariant format of the action. Restricting our study for illustration to electrostatics, where only $\vec E$ and $\vec D$ do not vanish, we can easily understand Born\rq s idea. Consider
\begin{equation}\label{BI1}
{\cal L}\equiv -\epsilon_0 E_\mathrm{BI}^2\left(\sqrt{1-\vec E^{\,2}/E_\mathrm{BI}^2}-1\right) \to 
\displaystyle\frac{\epsilon_0}{2} \vec E^{\,2}\;,
\end{equation} 
where the weak field limit is indicated. For the $\vec D$ field we obtain
\begin{equation}\label{BI2}
\vec D=\displaystyle \frac{\partial \cal{L}}{\partial \vec E}=
 \displaystyle \frac{\vec E}{\sqrt{1-\vec E^{\,2}/E_\mathrm{BI}^2}}\;,\qquad 
%\end{equation} 
%\begin{equation}\label{BI3}
\vec E= \displaystyle \frac{\vec D}{\sqrt{1+\vec D^{\,2}/E_\mathrm{BI}^2}}\;.
\end{equation} 
We see that when $|\vec E|\to E_\mathrm{BI}$ the displacement field $\vec D$ diverges. Thus for a solution of the Coulomb problem inherent in \req{LF2}, the field $\vec D\propto \hat r/r^2$ diverges at the origin for a point source, while at the origin the electrical field $\vec E$ reaches its maximum BI value. More generally and independent of the form of the displacement field $\vec D$, we find $\vec E^{\,2} \leq E_\mathrm{BI}^2$. 

Considering the Lorentz-force on a particle due to a point (nuclear) source
\begin{equation}\label{LF1}
%\displaystyle \frac{d\,p^\mu}{d\tau}=eF^{\mu\nu}u_nu\;,\quad +e\vec v\times \vec B
\frac{d\vec p}{dt}=e\vec E= 
 -\displaystyle \frac{\hat r\,eE_\mathrm{BI}}{\sqrt{r^4/(Z r_\mathrm{BI}^2)^2 +1}}\;,\qquad 
eE_\mathrm{BI}\equiv \displaystyle\frac{e^2}{4\pi\epsilon_0r_\mathrm{BI}^2}\;,
\end{equation}
we see that there is a limited strength force. The negative sign appears since the electron and the nucleus carry opposite charges. 

Born invented (and Born and Infeld improved) a limit to force and this is what attracted the attention of Walter Greiner. Seeing a limit to force one may justly ask if there is a limit on electron binding in the field of a heavy nucleus. However, a classical limit to electric force does not mean that the potential governing solutions of the Dirac equation is is also bounded. The electron atom is characterized by a weak electric field stretching far when compared to the radius of the K-shell electron inside a heavy atom. Therefore to limit the electron binding one must limit the potential depth. $|eV|<2mc^2$ is required to prevent the electron from \lq\lq diving\rq\rq\ into the negative continuum. 

To quantify this we evaluate the radial integral of \req{LF1}
\begin{equation}\label{VF1}
eV(r)= \int_r^\infty e E_rdr\;, \qquad
eV_\mathrm{BI}(0)=- \displaystyle\frac{1.8541\alpha Z^{1/2} \hbar c}{r_\mathrm{BI}}\;.
\end{equation}
We show the value at origin as $V(r)$ is a complicated hypergeometric function. The BI choice of $r_\mathrm{BI}$ was made such that the electron mass is accounted for as being the energy content of the electromagnetic field: $r_\mathrm{BI}=1.236r_e=3.483 $\,fm, where $r_e=e^2/(4\pi\epsilon_0 mc^2) $ is the classical electron radius and the numerical factor follows from some technicalities, see for example Chapter 28 in Ref.\,\cite{Rafelski:2017hyt}. 

We thus discover that 
\begin{equation}\label{VF1bis}
eV_\mathrm{BI}(0)=- \displaystyle\frac{ Z^{1/2}\,2.666 \mathrm{MeV\,fm}}{r_\mathrm{BI}}
= - Z^{1/2} 0.765\,\mathrm{MeV}\;.
\end{equation}
We see that the depth of the potential is finite but unbounded. $eV_\mathrm{BI}(0)$ scales with $Z^{1/2}$ instead of the $Z$ we are familiar with in the linear Maxwell theory. One can interpret this result in the context of Maxwell theory, introducing an atomic nucleus effective size $R=Z^{1/2}2.75$\,fm. This size is, however, much too large. In order to be compatible with atomic physics data, a much larger BI limiting field is required~\cite{Rafelski:1971xw,Rafelski:1971kta,Rafelski:1973fh,Rafelski:1973fm}. 

With a larger $E_\mathrm{BO}$, according to \req{LF1}, we would need a smaller $r_\mathrm{BI}$, perhaps $r_\mathrm{BI}\to r_\mathrm{BI}/5$. Since the EM mass of the electron scales with $1/r_\mathrm{BI}$, such a field implies an electron EM mass well in excess of experiment. To summarize, the conclusion is that there is on one hand no stability of atomic orbitals, and on the other, the key attractions of BI theory is invalidated: the EM field energy for an electron is clearly too large after the BI limiting field parameter is adjusted to agree with atomic experimental constraints. This was the result that made Walter search for the understanding of what happens when a Dirac state as a function of parameter such as nuclear charge $Z$ mutates from being an electron into new existence of a positron. In colloquial language we call this \lq\lq diving\rq\rq.

%%%%%%%%%%%%%%%%%%%%%%%%%%%%%%%%%%%%%%%%%%%%%%%%
\section{Positron Production and Charged Vacuum}\label{CharVac}
For an uninitiated reader the first necessary insight is why we call the Coulomb potential that is capable of binding an electron by more than $2m_ec^2$ \lq\lq supercritical\rq\rq. To answer this question, let us consider the electron-positron $e^-e^+$-pair production process. The minimum energy required is $2m_ec^2$. However, in the presence of a nucleus of charge $Ze$, it is possible that we do not require this vacuum energy, since there is an electronic bound to the nucleus, and the binding reduces the pair energy threshold. 

%%%%%%%%%%%%%%%%%%%%%%%%%%%%%%%%%%%%%%%%%%%%%%%%
\subsection{Dirac equation and \lq\lq diving\rq\rq}\label{FanoD}
The threshold for pair conversion of a $\gamma$-ray to an $e^-e^+$-pair in the presence of a nucleus is 
\begin{equation}
E_T^\gamma = m_ec^2+\epsilon_n\;,
\end{equation} 
where $\epsilon_n$ is the energy of the bound electron (always including its rest mass) in the eigenstate $n$. Considering the Pauli principle we recognize that this is only possible if such a state has not been occupied by another electron. The above energy balance for the $\gamma$-conversion to $e^-e^+$ pair implies the following statement:
\begin{quote} {\em When $\epsilon_n\to -m_ec^2$, the minimum energy required to create an $e^-e^+$-pair approaches zero: $E_T^\gamma\to 0$. At the critical point $\epsilon_n= -m_ec^2$, the energy of the ionized atom is equal to the energy of the atom with a filled $1s$-electron state and a free positron of nearly zero kinetic energy.}
\end{quote} 

It is important to consider carefully what happens if and when a metastable bound state $\epsilon_n\to \epsilon_R<-m_ec^2$ could exist. In such a situation the energy of a fully ionized atom without the $1s$-electron(s) is higher than the energy of an atom with a \lq\lq filled\rq\rq\ K-shell and free positron(s). Thus a bare supercritical atomic nucleus cannot be a stable ground state and therefore the neutral (speaking of electro-positron) vacuum cannot be a stable ground state either.

We conclude that for super-critical binding where a quasi-state dives into the negative energy sea as we see in \rf{fig:SQMDiving}, the supercritical bare atomic nucleus will spontaneously emit a positron $e^+$ (or two $e^+$, allowing for spin), keeping in its vicinity the accompanying negative charge which thus can be called the real vacuum polarization charge. The state that has an undressed atomic nucleus is the \lq\lq neutral vacuum\rq\rq\ (vacuum for electrons, positrons), and beyond the critical point is not the state of lowest energy. The new state of lower energy, called the charged vacuum~\cite{Rafelski:1974rh}, is the dressed atomic nucleus; that is a nucleus surrounded by the real vacuum polarization charge.
 
The physics understanding here described was created in early 1972~\cite{Muller:1972zza}; this was the great insight Walter gained one Saturday morning, based on the Fano resonance embedment method. However, when seen in hindsight we could arrive at this result solving the Dirac equation to determine phase shifts of positron scattering states, which we did a year after~\cite{Rafelski:1973fj}. The phase shift analysis shows resonant scattering and allows the determination of the width of the resonance reliably.

%%%%%%%%%%%%%%%%%%%%%%%%%%%%%%%%%%%%%%%%%%%%%%%%
\subsection{Quasi molecules}\label{QMol}
Critical binding requires a super-superheavy nucleus containing $Z\ge173$ protons within a realistic nuclear volume. On the other hand there is a fundamental interest in seeing the vacuum decay predicted. In 1971 we recognized that in heavy-ion collisions the relativistic deeply bound electrons were moving fast enough to form quasi-molecular states around the two slowly moving nuclear Coulomb potential centers. 

This means that the collision of two extremely heavy nuclei could be used to probe the supercritical fields; in the second paragraph in Ref.\,\cite{Rafelski:1971xw} we state: \lq\lq Even if superheavy elements cannot be readily produced, enough information could possibly be gathered in the collisions of heavy ions, such as Pb on Pb or Cf on Cf, to decide if this limit exists. In these collisions the adiabatic approximation should have some validity since the velocity of the electrons in the $1s$ and $2p$ atomic orbitals is much faster than the relative nuclear velocity. Hence, as far as the electrons in the lower atomic orbitals are concerned, the collisions of Pb on Pb and of Cf on Cf may simulate superheavy electronic molecules with $Z=164$ and $Z =196$, respectively.\rq\rq

The relatively slowly moving heavy-ions with energies chosen to stop the collision at the Coulomb barrier provide a common field for a shared quasi-molecular electron cloud. These electron eigenstates could be computed in a good approximation using the combined Coulomb field corresponding to a super-heavy nucleus of charge $2Z$, with a quasi potential formed by an effective nuclear charge distribution with a diameter $2R_N=R_{12}$, corresponding to the distance $R_{12}$ between the two nuclei~\cite{Rafelski:1972mf}. 

This \lq\lq monopole\rq\rq\ approximation can be justified by averaging the two lowest terms in the multipole expansion. Adopting such an effective radial form of the potential to simulate the effect of axially symmetric potential implements the idea of quasi-molecular states where the electrons circle around the two centers, or seen in reverse, the two nuclear charges circle around each other, and the electron is observing the so obtained averaged potential. The shape of the adopted effective monopole radial potential is
\begin{align}
V_0(r)=
\begin{cases}
-\frac{3}{2}\frac{Z\alpha}{(R_{12}/2)}\left(1-\frac{r^2}{3(R_{12}/2)^2}\right)\ &\mbox{for}\ \ 0\le r\le R_{12}/2\\[0.3cm]
-\frac{Z\alpha}{r} \ \ &\mbox{for}\ \ R_{12}/2 < r< \infty\;.
\end{cases}\label{eq:eqO-2-11}
\end{align}

In \rf{Fig_m12} the exact two center potential following the axis connecting the two nuclei (dashed line) is compared to the monopole approximate potential (solid line) for the case of a Uranium-Uranium collision. We show both potentials for $R_{12}\equiv R=38.6$\,fm, the critical separation between the two Uranium nuclei. This shows that the electrons experience attractive forces similar to those of a super-heavy nucleus with $Z_\mathrm{eff}=184$, protons.

%%%%%%%%%%%%%
\begin{figure}[t]%[htb]
\sidecaption%[t]
\includegraphics[width=0.64\columnwidth]{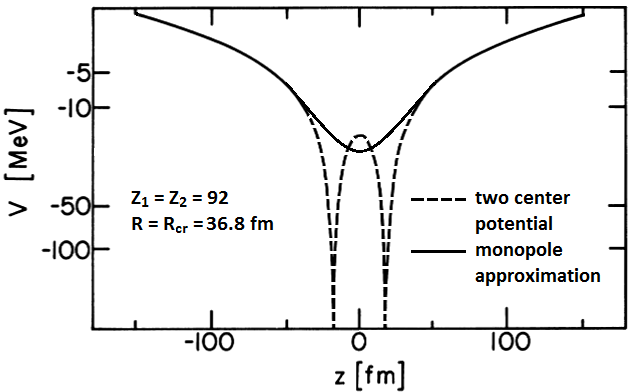}
%\centering
%\resizebox{0.7\textwidth}{!}{%
%\includegraphics{./Pics/Fig_m12.png}
%}
\caption[]{
Solid line: the (averaged) monopole potential that can be used to compute the electron binding in presence of colliding heavy-ions, see text; dashed line: the two center potential cut along the axis connecting the two nuclei. Source: Ref.\,\cite{Rafelski:2016ixr}}\label{Fig_m12}
\end{figure}
%%%%%%%%%%%%%

This simple approximation was tested extensively later, using the numerical methods that were developed in Ref.\,\cite{Rafelski:1975rf}, and found to be a very accurate and useful tool in understanding the physics of strong fields in heavy-ion collisions at sub- and near-Coulomb barrier collisions.

%%%%%%%%%%%%%%%%%%%%%%%%%%%%%%%%%%%%%%%%%%%%%%%%
\subsection{Experiments on positron production}
The following experimentally observable effect emerges as a consequence of the supercritical binding: in collisions of high $Z$ heavy ions an empty $1s$-state can be bound by more than $2m_ec^2$. Subsequently, a positron is emitted spontaneously. When the heavy ions separate again, the previously empty $1s$-state is now occupied by an electron; thus we effectively produced a pair by spontaneous vacuum decay. The theoretical treatment of the process is greatly facilitated by the large mass of the two nuclei: the Sommerfeld parameter $\eta=Z_1Z_2\alpha/v> 500$. Hence the classical approximation to the nuclear motion is adequate, and only the electrons have to be treated quantum mechanically.

The actual physical situation is not that simple: the heavy-ion collision is a time-dependent process; thus there may not always be enough time to emit a positron. Moreover there are several processes driven by time dependence of the collision, see \rf{Fig_m15}. For the positron production to involve the tightly bound eigenstate we need to remove electrons still present in the K-shell quasi-molecular states, see processes $a,b$. The motion of the ions can induce positron production in the processes $d,e$, there can be furthermore direct free pair production process $f$. Coherently superposed to processes $d,e,f$ is the spontaneous positron emission process $c$. Detailed discussion of the extensive 1970-1981 study of the theoretically anticipated effects can be found in~\cite{JoReinStick81}.

%%%%%%%%%%%%%%%%%%%%%%%%%%%%%%%%%
\begin{figure}[t]%[htb]
\sidecaption%[t]
\includegraphics[width=0.64\columnwidth]{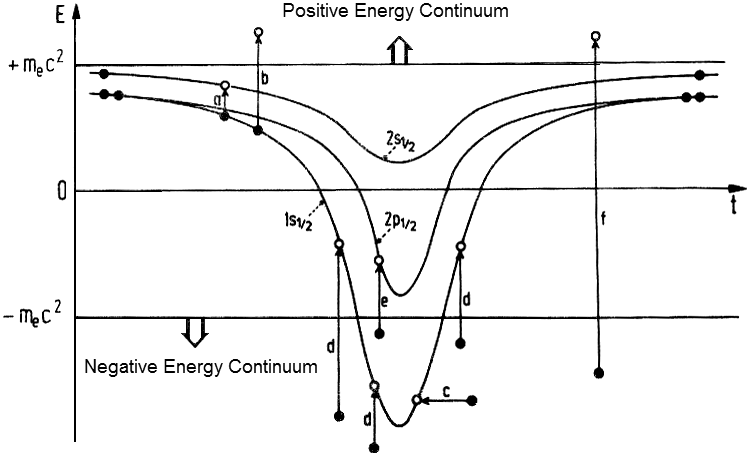}
%
%\centering{
%\resizebox{0.7\textwidth}{!}{%
%\includegraphics{./Pics/Fig_m15.png}
%}
%}
\caption[]{Schematic representation of pair-production processes
in heavy-ion collision as a function of time. We see most tightly bound eigenstates and relevant processes: $a,b$-ionization; $c$-spontaneous
and $d,e$-induced vacuum decay, $f$-continuum pair production. Source: Ref.\,\cite{Rafelski:2016ixr}}\label{Fig_m15}
\end{figure}
%%%%%%%%%%%%%%%%%%%%%%%%%%%%%%%%%

The rather short lifetime of a supercritical K-shell vacancy against positron emission, $\tau_{e^+}\simeq 10^{-18}$--$10^{-19}$\,sec implies that the supercritical system needs to live only for such a short period of time. An estimate of the order of magnitude of collision time shows that this is indeed feasible: the typical collision time of two nuclei at energies just below the Coulomb barrier is
\begin{equation}
\tau_{\rm coll}\simeq \displaystyle \frac{2R_\mathrm{cr}}{v}\simeq 0.25 \times 10^{-20}\mbox{sec}\;,\label{eq:eqO-3-1}
\end{equation}
with $R_\mathrm{cr}\simeq 35$\,fm (see below). The emission time for positrons is typically 100 times longer such that one expects a yield of roughly 1\% in this reaction.

The eigenstate energy of of most tightly bound electrons increases as ions approach and at $R_\mathrm{cr}\simeq 35$\,fm, it equals $-2m_e$ a for the $1s_{1/2}\sigma$ electron state in U+U collisions~\cite{Rafelski:1976mu}. The quasi-molecule is rendered supercritical in just the same way as the super-heavy atom was at $Z> Z_\mathrm{cr}$. 

A lot of effort went into the experimental search for spontaneously emitted positrons. A contemporary discussion of the experimental results has been recently presented~\cite{Rafelski:2016ixr}. In a nutshell, the consensus view today is that positrons observed were due to system dependent nuclear excitations converting into pairs. Walter Greiner was deeply marked by these disappointing experimental developments. In fact the word \lq\lq disappointment\rq\rq\ does not even come close to describing his feelings. While today there is no ongoing heavy ion positron production experiment, many regret that the experimental effort ended without an experimental result addressing strong field physics.

\section{Our life}
These comments about the life and work of my teacher are best concluded with a few pictures that tell more about the lasting relationship Walter Greiner enjoyed with the author and his family.
%%%%%%%%%%%%%%%%%%%%%%%%%%%%%%%%%
\begin{figure}%[t]%[htb]
\sidecaption%[t]
\includegraphics[width=0.64\columnwidth]{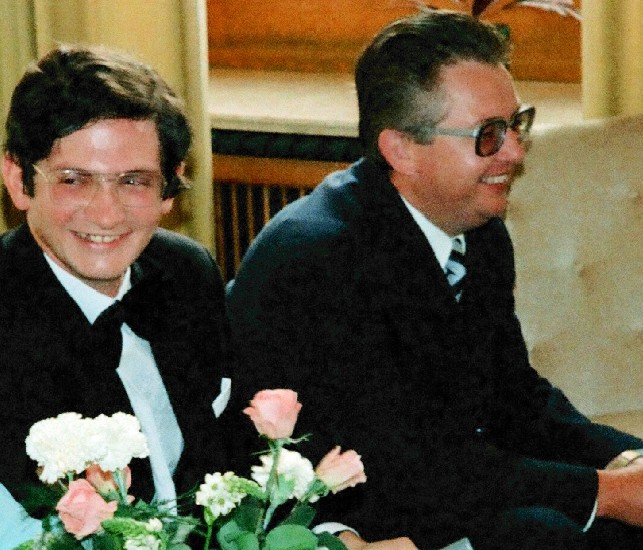}
\caption[]{Walter Greiner prepares to become witness at the marriage of Johann Rafelski (in picture) and Helga E. Betz, August 1973 at the Frankfurt R\"omer {\it Photo: Johann Rafelski}}\label{marriage}
\end{figure}
%%%%%%%%%%%%%%%%%%%%%%%%%%%%%%%%%

%%%%%%%%%%%%%%%%%%%%%%%%%%%%%%%%%
\begin{figure}%[t]%[htb]
\sidecaption%[t]
\includegraphics[width=0.64\columnwidth]{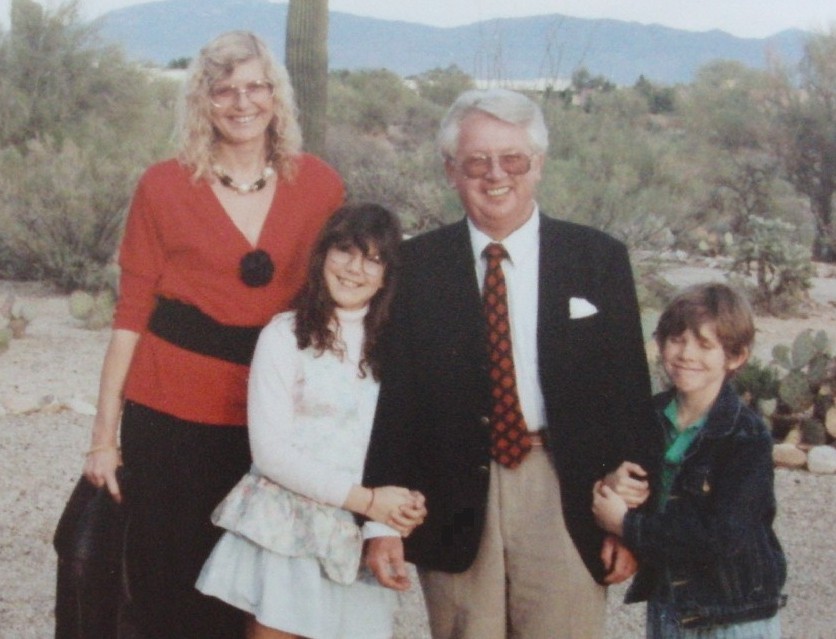}
\caption[]{Walter Greiner with the Rafelski Family in Tucson, 1988. {\it Photo: Johann Rafelski}}\label{Foothills}
\end{figure}
%%%%%%%%%%%%%%%%%%%%%%%%%%%%%%%%%

%%%%%%%%%%%%%%%%%%%%%%%%%%%%%%%%%
\begin{figure}%[t]%[htb]
\sidecaption%[t]
\includegraphics[width=0.64\columnwidth]{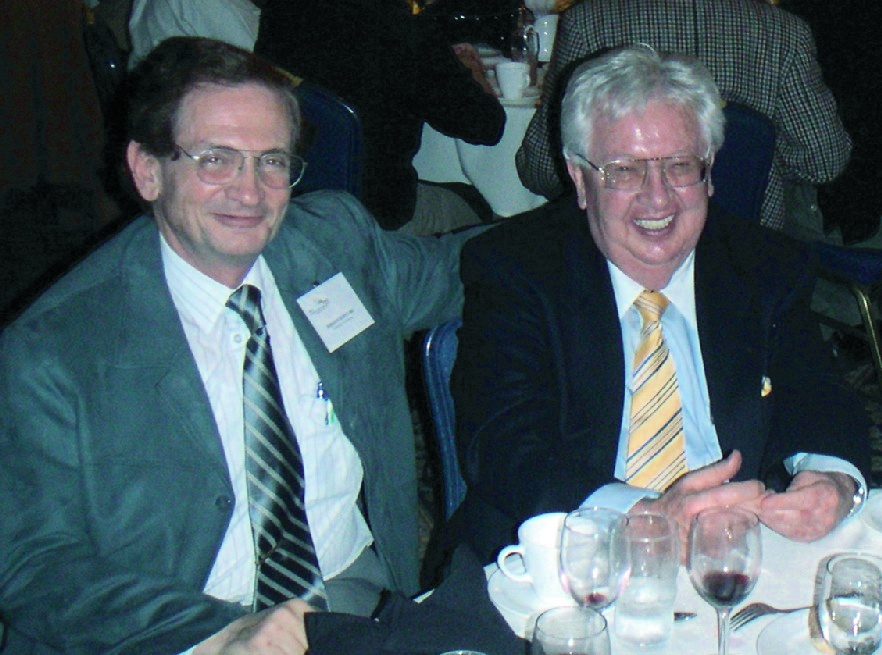}
\caption[]{Walter Greiner and the author in March 2006 Dinner at SQM2006, UCLA. {\it Photo: Johann Rafelski}}\label{SQM}
\end{figure}
%%%%%%%%%%%%%%%%%%%%%%%%%%%%%%%%%

\begin{acknowledgement}
The author thanks FIAS and Horst St\"ocker for support making this presentation possible. %Presented at FIAS International Symposium on Discoveries at the Frontiers of Science, Frankfurt, June 2017.
\end{acknowledgement}
%
%%%%%%%%%%%%%%%%%%%%%%%%%%%%%%%%%%%%%%%%%%%%%%%%%%%%%%%%%%%%%%%%%%%%%%%%%%%%%%%%%%%%%%%%%%

\end{document}